\documentclass[aps,twocolumn,showpacs,preprintnumbers,amsmath,amssymb]{revtex4}

\usepackage{graphicx}
\usepackage{dcolumn}
\usepackage{bm}
\usepackage{txfonts}
\usepackage{lscape}
\begin{document}

\title{Generation of $d$-wave coupling in the two-dimensional Hubbard\\ model from functional renormalization}

\author{H. C. Krahl${}^{a}$}  
\author{J. A. M\"uller${}^{a,b}$}  
\author{C. Wetterich${}^{a}$}

\affiliation{\mbox{\it ${}^a$Institut f{\"u}r Theoretische Physik,
Universit\"at Heidelberg,
Philosophenweg 16, D-69120 Heidelberg, Germany}\\
\mbox{\it ${}^b$Institut f\"ur Kernphysik,
Technische Universit\"at Darmstadt, Schlossgartenstra{\ss}e 9},\\
\mbox{\it D-64289 Darmstadt, Germany}}

\begin{abstract}
Within the two-dimensional repulsive $t-t'$-Hubbard model, an attractive coupling in the $d$-wave pairing channel is induced by antiferromagnetic fluctuations. We investigate this coupling using functional renormalization group equations. The momentum dependent $d$-wave coupling can be bosonized by the use of scale dependent field transformations. We propose an effective coarse grained model for the Hubbard model which is based on the exchange of antiferromagnetic and $d$-wave collective bosons. 
\end{abstract}



\maketitle
\section{Introduction}

The Hubbard model \cite{hubbard,kanamori,gutzwiller} for strongly correlated electrons has been used for a wide variety of phenomena, ranging from high temperature superconductivity \cite{anderson} to the metal insulator transition or antiferromagnetism. Solving this model is a major theoretical challenge. This is due to the complexity of the effective electron interaction, typically characterized by the competition of different channels, like the antiferromagnetic  or $d$-wave Cooper-pair exchange channels  \cite{scalapino,miyake,loh,bickers,bickersscalapinowhite,dahm,schmalian,millismonien,millisbalatsky}. So far, the development of $d$-wave superconductivity is still a controversial issue. Although several many-body techniques predict the emergence of a $d$-wave instability as leading instability in certain parameter ranges, see e. g. \cite{scalapino}, numerical studies have difficulties to detect superconductivity \cite{scalapino,moreo,assaad,dagotto,maier,veilleux,zhang,aimi}, as a consequence of both finite size and temperature limitations.

The $d$-wave-pairing interactions are absent in the microscopic Hubbard model.
As a plausible mechanism leading to $d$-wave pairing the exchange of effective  antiferromagnetic bosonic degrees of freedom has been proposed \cite{miyake,loh,bickers,bickersscalapinowhite,dahm,schmalian}. This idea was also suggested by \cite{millismonien,millisbalatsky} on the basis of a phenomenological spin-spin susceptibility. The generation of the interaction in the $d$-wave channel has been investigated extensively by the functional renormalization group study of a momentum dependent four electron vertex \cite{zanchi1,zanchi2,halbothmetzner,halbothmetzner2,salmhofer,honerkamp01,honerkampsalmhofer01,honerkamp02,katanin,metznerreissrohe,reissrohemetzner}. Our investigation employs non-perturbative flow equations, based on exact renormalization group equations for the average action or flowing action \cite{cw93}, \cite{berges_review02}. It is complementary to studies of the electron vertex by focussing explicitely on the role of the antiferromagnetic bosons, using the technique of partial bosonization during the flow \cite{GiesWett,PawlowskiBB}. We investigate the two-dimensional Hubbard model with nearest and next to nearest neighbor hopping.

The aim of this note is the derivation of an effective coarse grained model, valid at intermediate length scales $k^{-1}$ larger than the lattice distance $a$, but smaller than the scale of macroscopic physics. Our effective model will be based on a description of the electron interactions by the exchange of collective bosons. Near half-filling a model with exchange of antiferromagnetic boson has been used \cite{bbw04,bbw05} for a quantitative computation of effective antiferromagnetic order below the effective critical temperature $T_c$. It is advocated that in two dimensions the size of the ordered domains increases exponentially as the temperature decreases towards zero. For $T<T_c$ this size exceeds the size of a typical experimental probe. For all practical purposes the physics is then the same as for an ordered system. This includes the existence of long range fluctations associated to Goldstone bosons. For $T_c<T<T_{pc}$ short range antiferromagnetic order is found, with a typical domain size smaller than the experimental probe. Here $T_{pc}$ denotes the pseudo-critical temperature below which short range antiferromagnetic order sets in. In this paper we extend this approach by including the exchange of collective bosons consisting of Cooper-pairs in the $d$-wave channel. They are assumed to be a crucial ingredient for the understanding of the Hubbard model away from half filling. Their condensate would lead to superconductivity.

We have investigated earlier the phase transition to superconductivity in an effective $d$-wave exchange model \cite{kw07}. The phase transition is found to be of the Kosterlitz-Thouless type \cite{kosterlitzthouless73}, characterized by (modified) essential scaling above $T_c$, a jump of the superfluid density at $T_c$, and a gapless excitation with temperature dependent anomalous dimension below $T_c$. The present note constitutes a step for establishing such an effective model as a coarse grained version of the microscopic Hubbard model. This mapping is not yet complete, since we concentrate here only on the generation of the effective electron interaction in the $d$-wave channel and its bosonization, while a more detailed investigation of the propagators and interactions of the $d$-wave bosons is postponed to a subsequent publication. 

The partially bosonized model opens the door for a straightforward computation of the flow below the pseudo-critical temperature $T_{pc}$, where local order sets in and higher order boson interactions (e.g. corresponding to eight-fermion interactions) play an important role. The low temperature region is notoriously difficult for a purely fermionic description since the four fermion interaction diverges as the pseudo-critical temperature $T_{pc}$ is approached from above.
Also the important effect of higher fermionic operators can be taken into account most conveniently in a partially bosonized formulation \cite{bbw04,bbw05}.
In the partially bosonized description the divergence of the four fermion interaction is due to a boson ``mass term'' or ``gap'' changing from positive to negative values during the flow. A negative mass term indicates local order, since at a given coarse graining scale $k$ the effective potential has a minimum for a nonzero value of the boson field. If this order persists for $k$ reaching a macroscopic scale, the model exhibits effectively spontaneous symmetry breaking, associated in our model to antiferromagnetism or $d$-wave superconductivity.

On the other hand, the partially bosonized formulation introduces some bias into the renormalization group procedure by focussing on one or several boson exchange channels. This can be circumvented in the purely fermionic renormalization group flow.
However, this bias can be systematically reduced by the introduction of further bosonic fields, which yield additional resolution for the description of the interaction between the electrons.
The present approach constitutes a first step into this direction and is thus complementary to references \cite{zanchi1,zanchi2,halbothmetzner,halbothmetzner2,salmhofer,honerkamp01,honerkampsalmhofer01,honerkamp02,katanin,metznerreissrohe,reissrohemetzner}.

\section{Effective average action}

We use a functional integral framework and investigate the effective average action $\Gamma_k$ \cite{cw93,berges_review02}, which includes all quantum and thermal fluctuations with momenta $\mathbf{q}^2\gtrsim k^2$. This is realized by introducing an infrared cutoff $R_k$ which suppresses the fluctuations with $\mathbf q^2\lesssim k^2$. This cutoff will be removed at the end for $k\rightarrow 0$. For $k=0$ one recovers the usual effective action -- the generating functional for the one particle irreducible Green functions. For $k\rightarrow\infty$, or $k\rightarrow\Lambda$, with $\Lambda$ a suitable ultraviolet scale, the fluctuation effects are negligible and $\Gamma_\Lambda$ becomes the microscopic or classical action $S_\Lambda$. Therefore the scale dependent average action $\Gamma_k$ interpolates between the microscopic action $S_{\Lambda}$ and the full quantum effective action $\Gamma$,
\begin{eqnarray}
\lim_{k\rightarrow\Lambda}\Gamma_k\simeq S_{\Lambda}\,,\quad
\lim_{k\rightarrow 0}\Gamma_k=\Gamma
\,.\end{eqnarray}

The $k$-dependent flow of the average action obeys an exact renormalization group equation \cite{cw93},
\begin{eqnarray}\label{eq:flowequation}
\partial_k\Gamma_k[\chi]=\frac{1}{2}\mathrm{STr}\{[\Gamma^{(2)}_k[\chi]+R_k]^{-1}\partial_kR_k\}
\,,\end{eqnarray}
where the ``supertrace'' $\mathrm{STr}$ runs over field type, momentum and internal indices, and has an additional minus sign for fermionic entries. The functional differential equation (\ref{eq:flowequation}) involves the full inverse propagator $\Gamma^{(2)}_k[\chi]$ (second functional derivative of $\Gamma_k$), regulated by $R_k$.
Approximations to the solution of Eq. (\ref{eq:flowequation}) proceed by a truncation on the rhs with a suitable ansatz for the form of $\Gamma_k$.

On the microscopic level, the Hubbard model is defined as a purely fermionic model for electrons on a lattice.
However, it is well known that also bosonic degrees of freedom, e.g. Cooper-pairs, play an important role on larger length scales.
Hence, the question arises how the optimal description of the relevant degrees of freedom on all scales can be achieved within a functional renormalization group study. On the one hand, we want to accommodate the purely fermionic model at high momentum scales. On the other hand, we want to study phase transitions and critical behavior for the macroscopic physics. 
In the vicinity of the critical temperature the long range fluctuations are dominated by bosonic composite operators.
In order to exploit directly the importance of the collective bosonic degrees of freedom, we employ partial bosonization as motivated by a Hubbard-Stratonovich transformation \cite{hubbardtransf,stratonovich}. 
Spontaneous symmetry breaking can then be described simply by nonzero expectation values of suitable bosonic fields. In a purely fermionic description the four fermion coupling often diverges for temperatures below a pseudo-critical temperature $T_{pc}$. This indicates the onset of local order for $T<T_{pc}$. In the partially bosonized version this simply translates to a vanishing ``mass parameter'' or gap for the bosons, and does not constitute an obstacle for investigations at $T<T_{pc}$.

Antiferromagnetic order for the two-dimensional Hubbard model at half filling has been successfully described in this framework \cite{bbw04,bbw05}. One investigates a bosonic field $\mathbf{a}$ which represents an antiferromagnetic fermion bilinear
\begin{eqnarray}\label{eq:spinwave}
\tilde{\mathbf{a}}(X) =\psi^\dagger(X)\boldsymbol{\sigma}\psi(X)e^{i\Pi X}
\,.\end{eqnarray}
We use the Matsubara formalism with Euclidean time $\tau$ compactified on a torus with circumference $\beta=T^{-1}$. The Matsubara frequencies for the fermions are $\omega=(2n+1)\pi T$, $n\in \mathbb{Z}$. Bosonic fields obey periodic boundary conditions such that $\omega=2n\pi T$.
We use a compact notation $X=(\tau,\mathbf x)$, $Q=(\omega,\mathbf{q})$ and 
\begin{eqnarray}\label{eq:sumdefinition}
\sum\limits_X=\int\limits_0^\beta d\tau\sum\limits_{\mathbf{x}},\quad\sum\limits_Q=T\sum\limits_{n=-\infty}^\infty \int\limits_{-\pi}^\pi \frac{d^2q}{(2\pi)^2}\,,\nonumber\\
\delta(X-X')=\delta(\tau-\tau')\delta_{\mathbf{x},\mathbf{x'}}\,,\hspace{1.5cm}\nonumber\\
\delta(Q-Q')=\beta\delta_{n,n'}(2\pi)^2\delta^{(2)}(\mathbf{q}-\mathbf{q'})\,.\hspace{1cm}
\end{eqnarray}
All components of $X$ or $Q$ are measured in units of the lattice distance $\mathrm a$ or $\mathrm{a}^{-1}$.
The discreteness of the lattice is reflected by the $2\pi$-periodicity of the momenta $\mathbf{q}$. The momentum $\Pi$ in Eq. (\ref{eq:spinwave}) is given by
\begin{eqnarray}\label{eq:commensurate}
\Pi=(0,\pi,\pi)
\,.\end{eqnarray}

Antiferromagnetic order is indicated by a constant nonzero expectation value $\langle\mathbf a(X)\rangle=\mathbf a_0$. The simplest description employs a quartic effective potential for $\mathbf a$
\begin{eqnarray}\label{eq:Ueff}
U_{\mathbf a}[\mathbf{a}]&=&\bar{m}^2_{a}\alpha
+\frac{1}{2}\bar\lambda_a\alpha^2
\,,\end{eqnarray}
with $\alpha=\mathbf a^2/2$. We will see that a vanishing of $\bar m_a^2$ corresponds to a diverging effective four fermion coupling, while negative $\bar m_a^2$ leads to a minimum of $U_{\mathbf a}$ at $\alpha_0\neq 0$, and therefore indicates spontaneous symmetry breaking.
In the renormalization group treatment, $\bar m_a^2$ and $\bar\lambda_a$ become $k$-dependent running couplings. A situation with $\bar m_a^2(k)<0$ for $k_{SR}<k<k_{SSB}$, $\bar m_a^2(k<k_{SR})>0$ describes local order in domains with linear scale between $k^{-1}_{SSB}$ and $k^{-1}_{SR}$, while no global antiferromagnetism is present. Macroscopic antiferromagnetic order can be observed if the $k$-dependent location of the potential minimum, $\alpha_0(k)$, stays nonzero as $k^{-1}$ reaches the size of the experimental probe. For $\bar m_a^2<0$ a crucial ingredient for the determination of $\alpha_0$ is the quartic bosonic coupling $\bar\lambda_a$. In a purely fermionic language this corresponds to an eight fermion vertex and is quite difficult to access.

We start with a Yukawa like ansatz for the effective average action
\begin{eqnarray}\label{eq:simplesttruncation}
\Gamma_k[\chi]=\Gamma_{F,k}[\chi]+\Gamma_{\mathbf a,k}[\chi]+\Gamma_{F\mathbf a,k}[\chi]
\,.\end{eqnarray}
It describes fermion fields $\psi$ and the ``antiferromagnetic boson field'' $\mathbf{a}$, with $\chi=(\mathbf{a},\psi,\psi^*)$.
The fermionic kinetic term 
\begin{eqnarray}
\Gamma_{F,k}=\sum_{Q}\psi^{\dagger}(Q)P_F(Q)\psi(Q)
\end{eqnarray}
involves the inverse fermion propagator
\begin{eqnarray}
P_{F}(Q)=i\omega+\xi(\mathbf q)
\,,\end{eqnarray}
where
\begin{eqnarray}
\xi(\mathbf q)=-\mu-2t(\cos q_1 +\cos q_2)-4t' \cos q_1\cos q_2
\,.\end{eqnarray}
This is the classical inverse propagator for the Hubbard model with next neighbor hopping $t$ and diagonal hopping $t'$. The chemical potential is denoted by $\mu$. We have neither included self-energy corrections for the fermionic propagator, whose significance is still under debate \cite{zanchiself,honerkampself,arracheaself}, nor possible Pomeranchuk instabilities, see e. g. \cite{halbothmetzner}.
In the present paper we restrict ourselves to a parameter and scale range where the spin correlations are maximal at the commensurate antiferromagnetic wave vector (\ref{eq:commensurate}), e.g., we use $t'=0$ or $t'/t=0.05$.

The purely bosonic term is described by  a kinetic term and a local effective potential $U_{\mathbf a}$, cf. Eq. (\ref{eq:Ueff}),
\begin{eqnarray}
\Gamma_{a,k}
       =\frac{1}{2}\sum_{Q}\mathbf{a}^{T}(-Q)P_{a}(Q)\mathbf{a}(Q) 
                      +\sum_XU_{a,k}[\mathbf{a}]
\,.\end{eqnarray}
The kinetic term $P_a$ involves the $Q$-dependent part of the inverse antiferromagnetic propagator and is discussed in detail in the appendix.
The Yukawa like interaction term couples the bosonic field to the fermions,
\begin{eqnarray}
\Gamma_{F\mathbf a,k}\!
&=&\!-\bar h_a\sum_Q\mathbf{a}^T(-Q)\tilde{\mathbf{a}}(Q)\\
&=&\!-\bar h_a\!\sum_{K,Q,Q'}\delta(K+\Pi-Q+Q')
        \mathbf{a}(K)\cdot[\psi^{\dagger}(Q)\boldsymbol{\sigma}\psi(Q')]\nonumber
\,.\end{eqnarray}
This simplest truncation contains five $k$-dependent couplings, namely $\bar m_a^2$, $\bar\lambda_a$, $\bar h_a$, as well as the wave function renormalization $A_a$ and the shape parameter $D$ which parameterize $P_a(Q)$.

Inserting our truncation into the exact flow equation (\ref{eq:flowequation}) yields a coupled system of flow equations for $\bar m_a^2$, $\bar\lambda_a$, $\bar h_a$ and $P_a$. This is solved numerically. We start at the microscopic scale $k=\Lambda$ with initial conditions
\begin{eqnarray}\label{eq:initialcond}
\bar{m}_a^2|_{\Lambda}=U_m\,,\quad\bar{h}_a|_{\Lambda}=U_m\,,\quad\bar\lambda_a|_{\Lambda}=0\,,\quad P_a(Q)|_{\Lambda}=0
\,.\end{eqnarray}
Since for this choice the effective action is quadratic in $\mathbf a$, it is easy to solve for $\mathbf a$ as a functional of $\psi$, $\psi^*$.
Reinserting this solution into $\Gamma_\Lambda$ yields the well known microscopic action for the Hubbard model, with four fermion coupling $U$ given by $U=3U_m$. Equivalently, we can use $S_{\Lambda}=\Gamma_{\Lambda}$ in the defining functional integral and perform the Gaussian integration over the ``auxiliary'' field $\mathbf a$. This shows that our model is equivalent to the fermionic Hubbard model \cite{bbw00}.

The equivalence between the Hubbard model and our description of antiferromagnetic boson exchange is exact, but this particular form of partial bosonization is not unique. Due to the possibility of a Fierz reordering of the local four fermion interaction one could also start with a coupling both in the charge and the antiferromagnetic channel on initial scale or include even further channels \cite{bbw00}. Distributing the local four fermion action differently into the bosonic channels would alter the relation $U=3U_m$. Since the different possible versions of partial bosonization are all equivalent to the Hubbard model, one may question the reliability of our quantitative results based on $U_m=U/3$. In fact, mean field theory shows a strong dependence of the phase diagram on the choice of partial bosonization \cite{bbw00}. However, mean field theory neglects the effects of bosonic fluctuations and their inclusion substantially reduces the dependence of the results on the choice of bosonization \cite{jaeckelw03,bbw04}. (Without truncations all exact partial bosonizations should be exactly equivalent, such that the dependence on the choice can be used as a check of the validity of approximations.) In this paper, we concentrate on antiferromagnetic and $d$-wave fluctuations and do not include the charge channel. The inclusion of charge fluctuations within our bosonized language is in principle possible, and a first study \cite{bbw04} has indeed revealed that the results depend only weakly on the initial distribution of the four fermion interaction into the antiferromagnetic and charge density wave channels.

\begin{figure}[t]
\includegraphics[width=45mm,angle=0.]{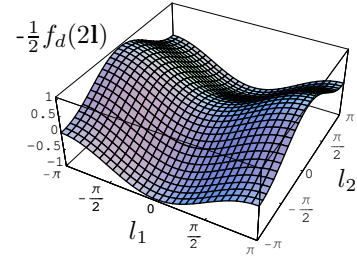}
\caption{\small{$d$-wave form factor.
}}
\label{fig:fd}
\end{figure}

In this paper we want to understand the physics associated to fermion bilinears whose expectation value describes $d$-wave superconductivity. We therefore have to extend the simplest truncation (\ref{eq:simplesttruncation}).
We introduce in addition a bosonic field $d$ associated to the appropriate Cooper-pairs $\tilde d$
\begin{eqnarray}\label{eq:dwave}
\!\!\tilde d(X)\!=\!\frac{1}{2}\psi^T\!(X\!-\!\frac{\hat e_1}{2})\epsilon\psi(X\!+\!\frac{\hat e_1}{2})\!-\!\frac{1}{2}\psi^T\!(X\!-\!\frac{\hat e_2}{2})\epsilon\psi(X\!+\!\frac{\hat e_2}{2})
.\end{eqnarray}
Here, $\hat e_1$, $\hat e_2$ are the unit vectors in the plane and $\epsilon=i\sigma_2$. (Eq. (\ref{eq:dwave}) is used here only as a shorthand for the definition of $\tilde d$ in momentum space, Eq. (\ref{eq:adtildemomspace}), since the Grassmann variables are not located at the lattice sites. A definition of $\tilde d$ in position space and the appropriate Fourier transformation can be found e.g. in \cite{scalapino}.)

For the purely bosonic part we add in our truncation solely a mass term of the $d$-boson
\begin{eqnarray}\label{eq:pured}
\Gamma_{d,k}=\bar m_d^2\sum_{Q} d^{*}(Q)d(Q)
\,.\end{eqnarray}
The $d$-field couples to the fermions by a Yukawa term
\begin{eqnarray}\label{eq:Fd}
\Gamma_{Fd,k}&=&-\bar h_d\sum_X(d^*(X)\tilde d(X) + d(X)\tilde d^*(X))
\\
&=&-\bar h_d\sum_Q(d^*(Q)\tilde d(Q) + d(Q)\tilde d^*(Q))\nonumber
\\
&=&-\frac{\bar h_d}{2}\sum_{K,Q,Q'}\delta(K\!-\!Q\!-\!Q')
    f_{d}(\mathbf q-\mathbf q')\nonumber\\
& &\hspace{0.cm}\times\Big(d^{*}(K)[\psi^{T}(Q)\epsilon \psi(Q')]
            -d(K)[\psi^{\dagger}(Q)\epsilon \psi^{*}(Q')] \Big)\nonumber
\,,\end{eqnarray}
with Yukawa coupling $\bar h_d$. The $d$-wave form factor
\begin{eqnarray}
f_d(\mathbf{q})\equiv\cos\frac{q_1}{2}-\cos\frac{q_2}{2}
\,,\end{eqnarray}
is kept fixed and shown in Fig. \ref{fig:fd}. We note the characteristic change of the sign under rotations of $90^\circ$. In our approximation, only the momentum independent coupling $\bar h_d$ depends on the scale $k$.
The extended truncation has two further running couplings, $\bar m_d^2$ and $\bar h_d$.

\begin{figure}[t]
\includegraphics[width=85mm,angle=0.]{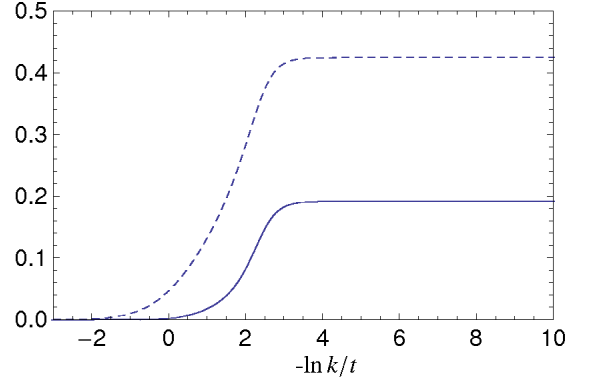}
\caption{\small{Generation of $d$-wave coupling. The solid line shows the flow of the $d$-wave channel fermionic coupling $\lambda_F^d$ for $T/t=0.13$, $\mu/t=-0.1$ and $t'=0$.  The dashed line shows the flow of the Yukawa coupling in the $d$-wave channel $\bar h_d$, as given by rebosonization for the same $T$, $\mu$, $t'$ and initial conditions (\ref{eq:initiald}).
}}
\label{fig:Yukawa_hd_flow}
\end{figure}

The initial values in the $d$-wave channel are
\begin{eqnarray}\label{eq:initiald}
\bar m_d^2|_{\Lambda}=1\,,\quad \bar h_d|_{\Lambda}=0
\,.\end{eqnarray}
At the microscopic scale the $d$-boson therefore decouples from the fermions and the $\mathbf a$-boson, such that the microscopic action is not modified.
Our model remains exactly equivalent to the Hubbard model. For $k<\Lambda$, however, a non-zero Yukawa coupling $\bar h_d$ is generated, as shown in Fig. \ref{fig:Yukawa_hd_flow}. ``Integrating out'' the $d$-boson by solving the field equations as a functional of $\psi$, and reinserting into $\Gamma_k$, yields now an effective four fermion interaction in the $d$-wave pairing channel
\begin{eqnarray}
\Gamma_{F,4}^d=\lambda_F^d\sum_X\tilde d^*(X)\tilde d(X)\,,\quad
\lambda_F^d=-\frac{\bar h^2_d}{\bar m^2_d}
\,.\end{eqnarray}
Even though absent microscopically, this interaction is generated during the flow by the coupling to the antiferromagnetic channel, as derived before in a purely fermionic language \cite{zanchi1,zanchi2,halbothmetzner,halbothmetzner2,honerkamp01,honerkamp02,honerkampsalmhofer01,katanin,metznerreissrohe,reissrohemetzner}. The effective coupling $\lambda_F^d$ is shown graphically in Fig. \ref{fig:Yukawa_hd_flow}.

\section{Flow Equations}

In addition to the truncation of the effective average action, the regulator functions have to be defined.
Our choice for the fermionic regulator is inspired by the fact that at nonzero temperature the inverse fermionic propagator \mbox{$P_F(Q)=2\pi i(n_F+\frac{1}{2})T+\xi_{\mathbf{q}}$} has no zero eigenvalue. The temperature itself acts as a regulator. We put this into use by the regulator function \cite{bbw04}
\begin{eqnarray}\label{eq:fermreg}
R^F_k(Q)=i\omega\big(\frac{T_k}{T}-1\big)=2\pi i(n_F+\frac{1}{2})(T_k-T)
\,,\end{eqnarray}
with
\begin{eqnarray}
T^4_k&=&T^4+k^4\,,\\
\partial_kT_k&=&(k/T_k)^3
\rightarrow\left\{\begin{array}{ll}
1 & \mathrm{if}\quad k\gg T\\
(k/T)^3 & \mathrm{if}\quad k\ll T
\end{array}\right.\nonumber
\,.\end{eqnarray}
For $k\gtrsim T$ the cutoff $R_k^F$ in the inverse fermion propagator
suppresses the contribution of all fluctuations with momenta
$|\mathbf{q}-\mathbf{q}_F|^2<(\pi k)^2$, even for $T=0$. It becomes
ineffective for $k\ll T$ where no cutoff is needed anyhow. Basically, the temperature $T$ is replaced by the scale dependent ``temperature'' $T_k$ -- we cool the fermions down to the temperature of interest during the flow. For $k\rightarrow\infty$ the cutoff diverges for all $T$ such that fermion fluctuations are completely suppressed and we start with the same initial conditions for all $T$. For $k\rightarrow 0$ the cutoff function vanishes for all $T$ and one recovers the quantum effective action. For finite $k^{-1}$, corresponding to a finite macroscopic size of the probe or a finite experimental wave length, the temperature dependence of the cutoff is small for $k\ll T$. In principal, different functional forms of $T_k(k)$ can be used for a test of the robustness of our truncation.

For bosons the situation is different. Here, long range bosonic modes may cause infrared problems which cannot be regularized by the same type of regulator function we use for the fermions. In particular, near a second order phase transition the bosonic correlation length diverges, which is the same as a vanishing boson mass term. Our regulator $R_k^a$ is devised in order to cut off the long range bosonic fluctuations. We specify the regularization of the $\mathbf a$-bosons in detail in the appendix. No regularization of the $d$-bosons is needed for the investigations in the present paper.

The flow equations for the couplings follow from projection of the flow equation (\ref{eq:flowequation}) onto the corresponding monomial of fields. For the mass parameter in the effective potential one derives \cite{bbw04}
\begin{eqnarray}\label{eq:flowm2a}
\partial_k\bar{m}_a^2&=&2\bar{h}_a^2\sum_Q\tilde\partial_k\frac{1}{P^k_F(Q)P^k_F(Q+\Pi)}\\
& &
+\frac{5}{2}\bar{\lambda}_a\sum_Q\tilde\partial_k\frac{1}{P_a^k(Q)+\bar{m}^2_a}\nonumber
\,,\end{eqnarray}
where
\begin{eqnarray}
P_a^k(Q)=P_a(Q)+R_k^a(Q)\,,\quad P_F^k(Q)=P_F(Q)+R_k^F(Q)
\,.\end{eqnarray}
The derivative $\tilde\partial_k$ acts only onto the explicit $k$-dependence introduced by the regulator functions, not on the couplings.
For the quartic coupling the flow equation reads
\begin{eqnarray}
\partial_k\bar{\lambda}_a&=&4\bar{h}_a^4\sum_Q\tilde\partial_k\frac{1}{(P^k_F(Q)P^k_F(Q+\Pi))^2}
\\
& &-
\frac{11}{2}\bar\lambda^2_a\sum_Q\tilde\partial_k\frac{1}{(P_a^k(Q)+\bar{m}^2_a)^2}\nonumber
\,.\end{eqnarray}
The flow of the kinetic term $P_a(Q)$ is discussed in the appendix.

The flow equation for the Yukawa coupling consists of a direct contribution $\beta^d_{\bar{h}_a}$ and a contribution $\beta^{rb}_{\bar{h}_a}$ which comes from rebosonization of regenerated fermionic couplings \cite{GiesWett},
\begin{eqnarray}
\partial_k\bar{h}_a=\beta^d_{\bar{h}_a}+\beta^{rb}_{\bar{h}_a}
\,.\end{eqnarray}
The first contribution reads
\begin{eqnarray}\label{eq:haflow}
\beta^d_{\bar{h}_a}=-\bar{h}_a^3\sum_Q\tilde\partial_k\frac{1}
{P_F^k(Q)P_F^k(Q+\Pi)(P_{a}^k(Q)+\bar{m}^2_a)}
\,.\end{eqnarray}
The second contribution is discussed in the following section, together with the flow equation for $\bar h_d$.
For the mass parameter $\bar m_d^2$ for the $d$-boson we find
\begin{eqnarray}
\partial_k\bar m^2_d=-\bar{h}_d^2\sum_Q\tilde\partial_k\frac{f_d^2(2\mathbf q)}{P_F^k(-Q)P_F^k(Q)}
\,.\end{eqnarray}

\section{Generation of $d$-wave coupling}

\begin{figure}[t]
\begin{center} 
\includegraphics[width=45mm]{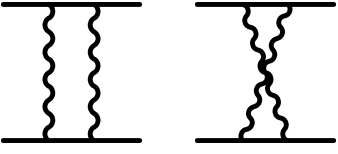}
\caption{\small{
One-loop corrections (box diagrams) to the four fermion interaction. Solid lines represent fermions, wiggly lines $\mathbf a$-bosons.
}}
\label{fig:fourfermion}
\end{center}
\end{figure}
In this section we discuss how a coupling in the $d$-wave channel is generated by the exchange of antiferromagnetic bosons.
This mechanism was suggested e.g. by \cite{scalapino,halbothmetzner}.
At the microscopic scale $\Lambda$ we have a local repulsive four fermion interaction, described in our picture by the exchange of $\mathbf a$-bosons. No other channel is present.
However, even in the case of purely repulsive forces between the electrons the system may become unstable against Cooper-pair formation \cite{kohnluttinger}.
On lower momentum scales we indeed observe a generation of an attractive coupling in a Cooper-pair channel which turns out to have $d$-wave symmetry. This may become critical in an appropriate parameter region.

We show in Fig. \ref{fig:fourfermion} the box diagrams exchanging $\mathbf{a}$-bosons. They generate an effective momentum dependent four fermion vertex $\Gamma_F^{(4)}$,
\begin{eqnarray}\label{eq:gamma4}
\Gamma_4\!\!&=&\!\!\frac{1}{4}\!\!\sum_{Q_1,\dots,Q_4}\Gamma^{(4)}_{F,\alpha\beta\gamma\delta}(Q_1,Q_2,Q_3,Q_4)\delta(Q_1-Q_2+Q_3-Q_4)\nonumber\\
& &\hspace{2cm}\times\psi^*_\alpha(Q_1)\psi_\beta(Q_2)\psi^*_\gamma(Q_3)\psi_\delta(Q_4)
\,.\end{eqnarray}
This irreducible vertex has to be added to the effective interaction arising from the exchange of bosons if we account for the total effective action between the electrons. It receives further contributions beyond the box diagrams since it has to include all contributions to the total four fermion vertex which are not accounted for by the antiferromagnetic boson exchange in a given truncation. In particular, the fluctuations of the antiferromagnetic bosons induce a momentum dependent piece in the effective Yukawa coupling between the electrons and the $\mathbf{a}$-bosons. This momentum dependent part is not reproduced by our truncation where a constant $\bar h_a$ is evaluated for a particular choice of external momenta. We have therefore added this piece to $\Gamma_4$. Our computation of the sum of $\Gamma_4$ and the antiferromagnetic boson exchange agrees with a purely fermionic computation in one loop order, such that the results are identical as long as $U/t$ remains small.

The $d$-wave channel interaction is part of $\Gamma^{(4)}(Q_i)$. Since the coupling $\Gamma^{(4)}(Q_i)$ contains contributions from various channels we will next project onto its contributions from the $d$-wave and the antiferromagnetic channels. For spin rotation invariant systems the spin structure of the four fermion interaction has the general form
\begin{eqnarray}
\!\!\!\!\Gamma^{(4)}_{F,\alpha\beta\gamma\delta}(Q_1,Q_2,Q_3,Q_4)&=&\Gamma^{(4)}_{F,s}(Q_1,Q_2,Q_3,Q_4)S_{\alpha\gamma;\beta\delta}\\
& &\hspace{-0.2cm}+\Gamma^{(4)}_{F,t}(Q_1,Q_2,Q_3,Q_4)T_{\alpha\gamma;\beta\delta}\nonumber
\,,\end{eqnarray}
where $S_{\alpha\gamma;\beta\delta}=\delta_{\alpha\beta}\delta_{\gamma\delta}-\delta_{\gamma\beta}\delta_{\alpha\delta}$ and $T_{\alpha\gamma;\beta\delta}=\delta_{\alpha\beta}\delta_{\gamma\delta}+\delta_{\gamma\beta}\delta_{\alpha\delta}$ project onto spin singlet and spin triplet states in the fermion pair channels, respectively.
Here, we are interested in spin singlet states and we choose $Q_1=-Q_3\equiv L=(\pi T,\mathbf l)$ and $Q_2=-Q_4\equiv L'=(\pi T,\mathbf l')$. For the projection onto the $d$-wave Cooper-pair channel we use the fact that the $d$-wave coupling changes its sign under rotation of $90^\circ$. We therefore define a momentum dependent $d$-wave channel coupling $\lambda_F^d(\mathbf l,\mathbf l')$ by
\begin{eqnarray}\label{eq:flowlamnda_d}
\lambda^d_F(\mathbf l,\mathbf l')
&=&-\frac{1}{2}\big\{\Gamma^{(4)}_{F,s}(L,L',-L,-L')
\\
& &\hspace{0.4cm}
-\Gamma^{(4)}_{F,s}(R(L),L',-R(L),-L')\big\}\nonumber
\,,\end{eqnarray}
where $R(L)$ denotes a rotation of the spatial components $\mathbf{l}$ of $L$ by $90^\circ$. For the definition (\ref{eq:flowlamnda_d}) we have kept $L'$ fixed and subtracted the same contribution after a rotation of $90^\circ$ of the space-like components of $L$. For instance, the antiferromagnetic channel and the $s$-wave channel are subtracted in this way.

\begin{figure}[t]

\includegraphics[width=85mm,angle=0.]{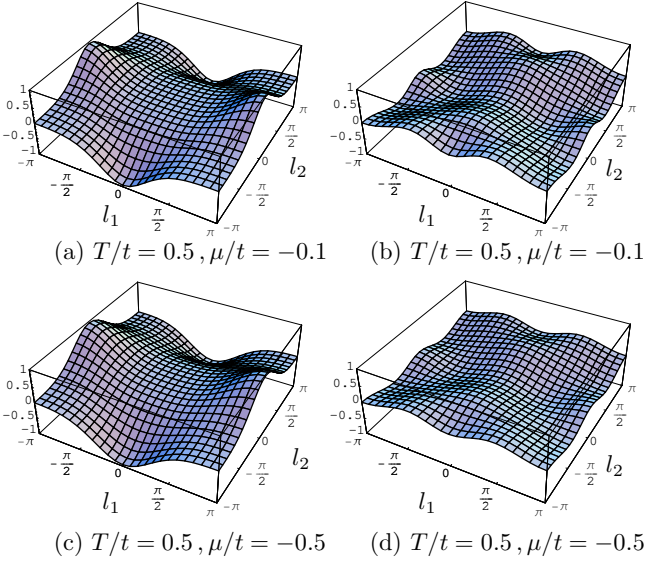}
\caption{\small{
Momentum dependence of $d$-wave coupling. In (a), (c) we show the normalized momentum dependence of the rhs of the flow equation for the fermionic coupling, $\frac{1}{4}\big(\partial_k\lambda^d_F(\mathbf l,\mathbf l')/\partial_k\lambda^d_F\big)$. In (b), (d) we display the residual coupling after the subtraction of the bosonized part, $\frac{1}{4}\big(\partial_k\lambda^d_F(\mathbf l,\mathbf l')/\partial_k\lambda^d_F-f_d(2\mathbf l)f_d(2\mathbf l')\big)$ (cf. Eq. (\ref{eq:flowlamnda_dsingle})). We use $t'=0$ in all plots. Notice that these plots are independent of $U$.}}
\label{fig:d-waveformfactor}
\end{figure}

We compute $\partial_k\Gamma_F^{(4)}$ and $\partial_k\lambda_F^d(\mathbf l,\mathbf l')$ by taking the fourth functional derivative of Eq. (\ref{eq:flowequation}) with respect to the fermion fields. Within our truncation this corresponds to the box diagrams in Fig. \ref{fig:fourfermion}. We find that $\partial_k\lambda_F^d(\mathbf l,\mathbf l')$ is well approximated by the flow of a simple $d$-wave channel coupling $\lambda_F^d$ of the form
\begin{eqnarray}\label{eq:flowlamnda_dsingle}
\lambda_F^d(\mathbf l,\mathbf l')=f_d(2\mathbf l)f_d(2\mathbf l')\lambda_F^d
\,.\end{eqnarray}
In Figs. \ref{fig:d-waveformfactor}a,c we show the flow of the lhs of Eq. (\ref{eq:flowlamnda_dsingle}) as a function of $\mathbf l$ for $\mathbf l'=(0,\pi)$ (for $k=\Lambda$ and with appropriate overall normalization).
There is a good qualitative agreement with the $d$-wave form factor displayed in Fig. \ref{fig:fd}. We determine the flow of $\lambda_F^d$ by choosing $\mathbf l=(\pi,0)$, $\mathbf l'=(0,\pi)$, i.e. $\partial_k\lambda_F^d=\partial_k\lambda_F^d((\pi,0),(0,\pi))$.

Next, we show how this fermionic quartic coupling can be bosonized on all scales $k$. This means that we trade the four fermion coupling $\lambda_F^d$ in favour of a Yukawa-coupling of the $d$-bosons. After this transformation, the $d$-boson exchange induces the effective interaction in the $d$-wave pairing channel. We achieve this reformulation by a variable transformation in the functional flow equation (\ref{eq:flowequation}). The flow equation (\ref{eq:flowequation}) describes the scale dependence of $\Gamma_k$ at fixed fields $\chi$. We will now switch to new $k$-dependent variables $\chi_k$ and derive new flow equations which describe the $k$-dependence of $\Gamma_k$ at fixed $\chi_k$. For a suitable choice of $\chi_k$ this will realize the desired bosonization \cite{GiesWett,PawlowskiBB}.

More precisely, we choose scale dependent bosonic fields $\mathbf B_k=\mathbf B_k[\psi,\psi^*,\mathbf B;k]=(\mathbf{a}_k,d_k,d_k^*)$
and perform the variable transformation in the exact flow equation
\begin{eqnarray}\label{eq:changegammarebos}
\frac{d}{dk}\Gamma_k[\psi,\psi^*,\mathbf B_k]\big{|}_{\mathbf B_k}
&=&\partial_k\Gamma_k[\psi,\psi^*,\mathbf B]\big{|}_{\mathbf B=\mathbf B_k}\\
& &+\sum_Q\Big(\frac{\delta}{\delta\mathbf{a}_k(Q)}
\Gamma_k[\psi,\psi^*,\mathbf B_k]\Big)
\partial_k\mathbf{a}_k(Q)\nonumber\\
& &+\sum_Q\Big(\frac{\delta}{\delta d_k(Q)}
\Gamma_k[\psi,\psi^*,\mathbf B_k]\Big)
\partial_k d_k(Q)\nonumber\\
& &+\sum_Q\Big(\frac{\delta}{\delta d^*_k(Q)}
\Gamma_k[\psi,\psi^*,\mathbf B_k]\Big)
\partial_k  d^*_k(Q)\nonumber
\,.\end{eqnarray}
This equation is our starting point for the construction of the ``perfect bosons'' \cite{GiesWett} on all scales.
The first term on the rhs is given by Eq. (\ref{eq:flowequation}).

The scale dependent boson fields are defined by adding suitable fermion bilinears
\begin{eqnarray}\label{eq:defalpha}
\mathbf a_k(Q)=\mathbf a_\Lambda(Q)-\alpha_k^a\tilde{\mathbf a}(Q)\,,\\
d_k(Q)=d_\Lambda(Q)-\alpha_k^d\tilde d(Q)\nonumber
\,.\end{eqnarray}
Here the fermion bilinears
\begin{eqnarray}\label{eq:adtildemomspace}
\tilde{\mathbf{a}}(Q)&=&\sum_P\psi^\dagger(P)\boldsymbol{\sigma}\psi(P-\Pi+Q)\,,\\
\tilde d(Q)&=&\frac{1}{2}\sum_Pf_d(2\mathbf p-\mathbf q)\psi^T(P)\epsilon\psi(Q-P)\nonumber
\,,\end{eqnarray}
correspond to the Fourier transforms of Eqs. (\ref{eq:spinwave}), (\ref{eq:dwave}).
In Eq. (\ref{eq:defalpha}), we denote by $\mathbf a_\Lambda$ and $d_\Lambda$ the original microscopic fields appearing in the microscopic action $\Gamma_\Lambda$. These are the $k$-independent fields which are kept fixed in Eq. (\ref{eq:flowequation}) or in the first term on the rhs of Eq. (\ref{eq:changegammarebos}). The fermion fields are kept scale independent and we infer
\begin{eqnarray}
\partial_k\mathbf{a}_k(Q)&=&-\partial_k\alpha_k^{a}\tilde{\mathbf{a}}(Q)
\,,\\
\partial_k d^{(*)}_k(Q)&=&-\partial_k\alpha_k^{d}\tilde d^{(*)}(Q)\nonumber
\,.\end{eqnarray}
\begin{figure}[t]
\includegraphics[width=50mm,angle=0.]{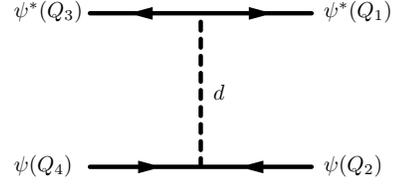}
\caption{\small{Effective four fermion interaction in the $d$-wave pairing channel mediated by $d$-boson exchange.}}
\label{fig:dexchange}
\end{figure}

The variable transformation in Eq. (\ref{eq:changegammarebos}) induces new Yukawa and four fermion couplings, given in our truncation by 
\begin{eqnarray}
\frac{d}{dk}\Gamma_k\big{|}_{\mathbf B_k}&=&\nonumber
\partial_k\Gamma_k\big{|}_{\mathbf B=\mathbf B_k}-\sum_{Q}(\partial_k\alpha_k^{a})(P_{a}(Q)+\bar m_a^2)\mathbf{a}_k(-Q)\cdot\tilde{\mathbf{a}}(Q)\\
& &-\sum_{Q}(\partial_k\alpha_k^{d})\bar m_d^2\big\{ d^*_k(Q)\tilde{d}(Q) + d_k(Q)\tilde{d}^*(Q)\big\}
 \nonumber\\
& &+\sum_{Q}(\partial_k\alpha_k^{a})\bar{h}_a\tilde{\mathbf{a}}(Q)\cdot
\tilde{\mathbf{a}}(-Q)\nonumber\\
& &+2\sum_{Q}(\partial_k\alpha_k^{d})\bar{h}_d\tilde{d}^*(Q)\tilde{d}(Q)
\,.\end{eqnarray}
Here the couplings $\bar h_a$, $\bar h_d$ etc. correspond to suitable functional derivatives of $\Gamma_k$ at fixed $\mathbf B_k$.

For a suitable choice of $\alpha_k^d$, we can use the four fermion interaction in the $d$-wave pairing channel generated by the variable transformation, in order to cancel a corresponding piece $\partial_k\lambda_F^d$ generated by the diagrams of Fig. \ref{fig:fourfermion} in $\partial_k\Gamma_k\big{|}_{\mathbf B}$. More precisely, we fix $\partial_k\alpha_k^d$ by the requirement
\begin{eqnarray}\label{eq:fixalphad}
0=\partial_k\lambda_F^{d}+2\bar{h}_{d}\partial_k\alpha^d_k
\,,\end{eqnarray}
where the first term is computed from Eqs. (\ref{eq:flowequation}), (\ref{eq:gamma4})-(\ref{eq:flowlamnda_dsingle}). The coupling $\lambda_F^d\big{|}_{\mathbf B_k}(\mathbf l,\mathbf l')$ is defined similar to Eq. (\ref{eq:flowlamnda_d}). However, it is now given by the functional derivative of $\Gamma_k$ at fixed $\mathbf B_k$.
This means that its flow is obtained from the flow of $\lambda_F^d(\mathbf l,\mathbf l')$ (at fixed $\mathbf B$) by subtracting the ``bosonized piece''
\begin{eqnarray}
\partial_k\lambda_F^d(\mathbf l,\mathbf l')\big{|}_{\mathbf B_k}=\partial_k\lambda_F^d(\mathbf l,\mathbf l')-\partial_k\lambda_F^df_d(2\mathbf l)f_d(2\mathbf l')
\,.\end{eqnarray}
In particular, if we define $\lambda_F^d\big{|}_{\mathbf B_k}$ in analogy to Eq. (\ref{eq:flowlamnda_dsingle}), one has $\partial_k\lambda_F^d\big{|}_{\mathbf B_k}=0$. We show $\partial_k\lambda_F^d(\mathbf l,\mathbf l')\big{|}_{\mathbf B_k}$ in Figs. \ref{fig:d-waveformfactor}b,d (in a suitable normalization where $\partial_k\lambda_F^d$ is divided out). Comparison with Figs. \ref{fig:d-waveformfactor}a,c shows that the major part of $\partial_k\lambda_F^d(\mathbf l,\mathbf l')$ can indeed be absorbed by the bosonization. The ``residual part'' $\partial_k\lambda_F^d(\mathbf l,\mathbf l')\big{|}_{\mathbf B_k}$ contains higher harmonics -- the dominant piece in Figs. \ref{fig:d-waveformfactor}b,d has a periodicity in momenta with period $\pi$, instead of the leading piece with periodicity $2\pi$ in Figs. \ref{fig:d-waveformfactor}a,c. In principle, this residual interaction can be partially absorbed by a more complex $k$-dependent field in Eq. (\ref{eq:defalpha}). We will not do this here and simply neglect in our truncation the residual interaction in $\partial_k\lambda_F^d(\mathbf l,\mathbf l')\big{|}_{\mathbf B_k}$. As can be seen by a comparison of Figs. \ref{fig:d-waveformfactor}b and \ref{fig:d-waveformfactor}d, the accuracy of this approximation gets better for larger $|\mu/t|$.
After the variable transformation, we can work with a vanishing four fermion coupling in the $d$-wave pairing channel.

From Eq. (\ref{eq:changegammarebos}) we can now infer the additional contributions to the flow of the Yukawa coupling $\bar h_d$. It accounts for the fact that after the variable change a leading contribution to the four fermion interaction is mediated by the exchange of bosons and no longer contained in $\lambda_F$. One obtains a modified flow equation for the Yukawa coupling for the $d$-boson
\begin{eqnarray}\label{eq:403}
\partial_k\bar{h}_{d}&=&\partial_k\bar{h}_{d}\big{|}_{\mathbf B}
+\bar m_d^2\partial_k\alpha^d_k
\,.\end{eqnarray}
Thereby inserting $\partial_k\alpha_k^d$ from Eq. (\ref{eq:fixalphad}) yields
\begin{eqnarray}\label{eq:Yuwawadflow}
\partial_k\bar{h}_{d}&=&\partial_k\bar{h}_{d}\big{|}_{\mathbf B}
-\frac{\bar m_d^2\partial_k\lambda_F^{d}}{2\bar{h}_{d}}
\,,\end{eqnarray}
for the $d$-boson. In our truncation one has $\partial_k\bar{h}_{d}\big{|}_{\mathbf B}=0$. The numerical solution to this flow equation is shown in Fig.  \ref{fig:Yukawa_hd_flow}.
We can reconstruct the effective four fermion interaction in the $d$-wave pairing channel by computing the tree diagram from the $d$-boson exchange (cf. Fig. \ref{fig:dexchange})
\begin{eqnarray}\label{eq:dexchange}
\lambda_{F,\mathrm{eff}}^d(Q_1,Q_2,Q_3,Q_4)&=&-\frac{\bar h_d^2}{\bar m_d^2}f_d(\mathbf{q}_1-\mathbf q_3)f_d(\mathbf q_2-\mathbf q_4)\\
& &\hspace{0.5cm}\times\delta(Q_1-Q_2+Q_3-Q_4)\nonumber
\,.\end{eqnarray}

\begin{figure}[t]
\includegraphics[width=80mm,angle=0.]{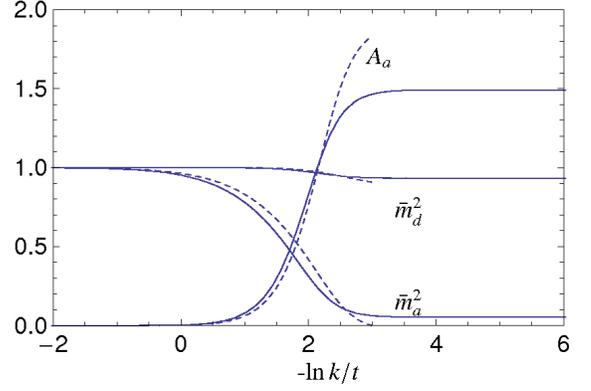}
\caption{\small{Scale dependence of $\bar m_a^2$, $\bar m_d^2$ and $A_a$ for temperature above $T_{pc}$ ($T/t=0.12$, solid lines) and below $T_{pc}$ ($T/t=0.09$, dashed lines). We use $\mu/t=-0.1$, $t'=0$, $U/t=3$.
}}
\label{fig:massesAa}
\end{figure}

\begin{figure}[t]
\hspace{-0.cm}\includegraphics[width=90mm,angle=0.]{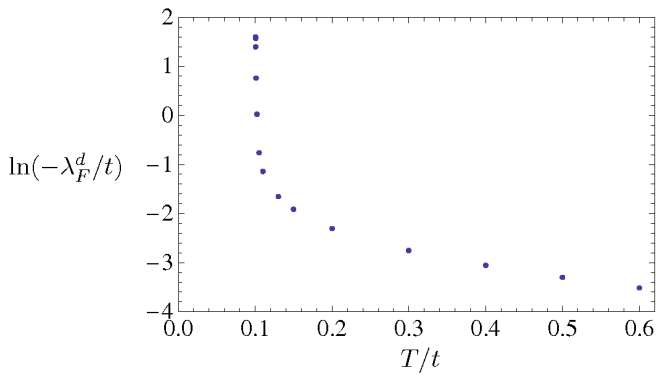}
\caption{\small{$\ln(-\lambda_F^d/t)$ for $k=0$ is plotted as a function of $T/t$ for $T\geq T_{pc}=0.1004t$, $\mu/t=-0.1$, $t'=0$.}}
\label{fig:lambdaFd_T}
\end{figure}

\begin{figure}[t]
\hspace{-0.cm}\includegraphics[width=90mm,angle=0.]{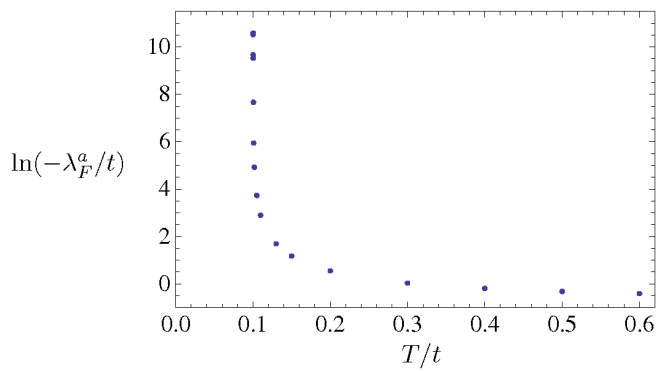}
\caption{\small{$\ln(-\lambda_F^a/t)$ for $k=0$ is plotted as a function of $T/t$ for $T\geq T_{pc}=0.1004t$, $\mu/t=-0.1$, $t'=0$.}}
\label{fig:lambdaFa_T}
\end{figure}

Similar as for the $d$-boson, the Yukawa coupling of the $\mathbf a$-boson receives an additional contribution due to the rebosonization of the four fermion interaction in the antiferromagnetic channel \cite{bbw04},
\begin{eqnarray}\label{eq:402}
\partial_k\bar{h}_{a}&=&\partial_k\bar{h}_{a}\big{|}_{\mathbf B}
+\bar m_a^2\partial_k\alpha^{a}_k
\,.\end{eqnarray}
We define the projection onto the four fermion interaction in the antiferromagnetic channel by
\begin{eqnarray}
\lambda_F^a=\frac{1}{8}\Gamma_{F,1221}^{(4)}(0,\Pi,0,-\Pi)
\,,\end{eqnarray}
and determine $\partial_k\alpha_k^a$  by
\begin{eqnarray}
0=\partial_k\lambda_F^a+\bar h_a\partial_k\alpha_k^a
\,.\end{eqnarray}
This yields
\begin{eqnarray}\label{eq:Yuwawaaflow}
\partial_k\bar{h}_{a}=\partial_k\bar{h}_{a}\big{|}_{\mathbf B}
-\frac{\bar m_a^2}{\bar{h}_{a}}
\partial_k\lambda_F^a=\beta^d_{\bar h_a}+\beta^{rb}_{\bar h_a}
\,,\end{eqnarray}
for the flow of the Yukawa coupling of the $\mathbf{a}$-boson. The first contribution was given in Eq. (\ref{eq:haflow}). The second contribution comes from rebosonization and we find in agreement with \cite{bbw04}
\begin{eqnarray}
\beta^{rb}_{\bar{h}_a}&=&\bar{m}_a^2\sum_{Q}\bar{h}_a^3\tilde\partial_k
\frac{1}{(P_a^k(Q)+\bar{m}_a^2)(P_a^k(\Pi-Q)+\bar{m}_a^2)}
\\
& &\hspace{1.5cm}\times\frac{1}{P_F^k(-Q)}\bigg\{\frac{1}{P_F^k(\Pi-Q)}-\frac{1}{P_F^k(Q)}\bigg\}\nonumber
\,.\end{eqnarray}
We have solved the flow equations for the couplings $\bar h_a$, $\bar h_d$, $\bar m^2_a$, $\bar m^2_d$, $A_a$ and $D$ numerically. In Fig. \ref{fig:massesAa} we show the $k$-dependence of $\bar m_a^2$, $\bar m_d^2$ and $A_a$. For $T>T_{pc}$ both $\bar m_a^2$ and $\bar m_d^2$ remain positive for all $k$.
We show the temperature dependence of the effective four fermion couplings in the $d$-wave and the antiferromagnetic channel, $\lambda_F^d$ and $\lambda_F^a$, in Figs. \ref{fig:lambdaFd_T} and \ref{fig:lambdaFa_T}. Both couplings are evaluated for $k=0$.

\section{Pseudocritical temperature}

\begin{figure}[t]
\includegraphics[width=80mm,angle=0.]{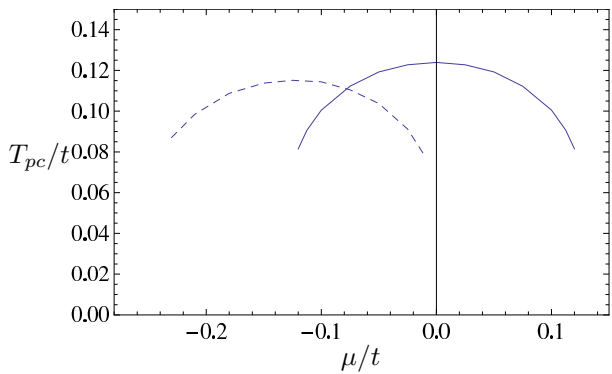}
\caption{\small{Pseudo-critical temperature $T_{pc}/t$ as a function of $\mu/t$ for $U/t=3$ and $t'=0$ (solid line) and $t'=-0.05t$ (dashed line).}}
\label{fig:phasediagronlyantif}
\end{figure}

\begin{figure}[t]
\includegraphics[width=80mm,angle=0.]{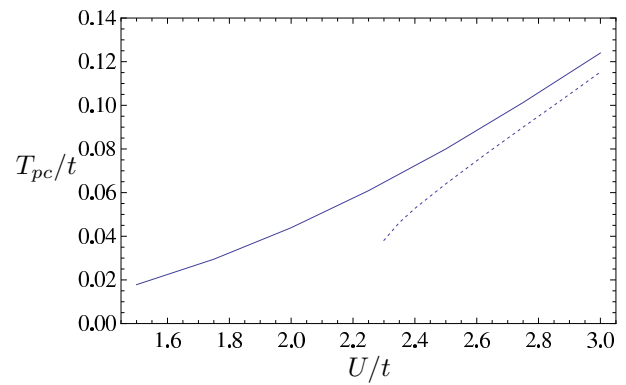}
\caption{\small{Pseudo-critical temperature $T_{pc}/t$ as a function of $U/t$ for $\mu=0\,,t'=0$ (solid line) and $\mu/t=-0.12\,,t'/t=-0.05$ (dashed line).}}
\label{fig:UdepofTpc}
\end{figure}

For $T\rightarrow T_{pc}$ the ``mass term'' $\bar m_a^2$ reaches zero at a nonzero value of $k$, and therefore $\lambda_F^a$ diverges. This feature extends to all $T\leq T_{pc}$ and signals the onset of local antiferromagnetic order in domains of linear size $k^{-1}$. For smaller $k$, the flow has to be continued in the ``spontaneously broken regime'' \cite{bbw04,bbw05}, where the minimum of the effective potential occurs for nonzero $\mathbf a$. In the present paper we do not continue the flow beyond the scale where $\bar m_a^2$ vanishes, but rather stop once $\bar m_a^2$ reaches zero. The corresponding pseudocritical temperature $T_{pc}$ is shown in Fig. \ref{fig:phasediagronlyantif} as a function of the chemical potential. We also show the dependence of $T_{pc}$ on the Hubbard coupling $U$ in Fig. \ref{fig:UdepofTpc}. For $U/t=3$ and $\mu=t'=0$ we find for $T_{pc}$ a value that is about $30\%$ smaller than in \cite{bbw04}. This is due to the different ansatz for the propagator of the $\mathbf a$-boson and may be taken as an indication for the size of the error.
In the present work we restrict ourselves to Hubbard couplings $U/t\leq 3$ and to temperature $T/t\gtrsim 0.02$. In order to extend the lines in Fig. \ref{fig:UdepofTpc} to smaller $T$ one has to improve our simple truncation of the frequency dependence of the propagator of the $\mathbf a$-boson. This issue will be addressed in a forthcoming work.

\begin{figure}[b]
\includegraphics[width=85mm,angle=0.]{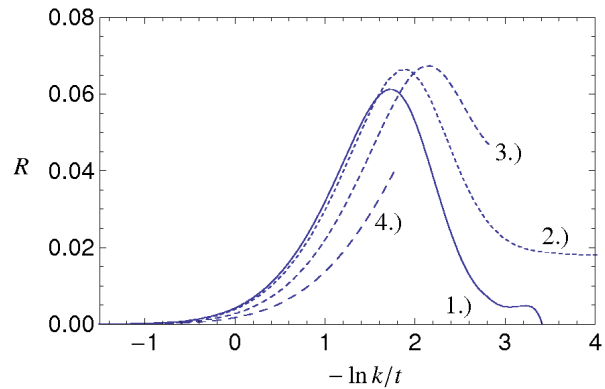}
\caption{\small{Relative strength of $d$-wave and antiferromagnetic coupling. The ratio $R$, defined in Eq. (\ref{eq:R}) is plotted for four parameter choices: 1.) $T/t=0.1235,\mu=0$, 2.) $T/t=0.115,\mu/t=-0.09$, 3.) $T/t=0.08,\mu/t=-0.15$ and 4.) $T/t=0.05,\mu/t=-0.3$. We use $U/t=3$, $t'=0$.}}
\label{fig:R}
\end{figure}

Another interesting quantity is the relative strength of the effective four fermion interaction in the $d$-wave pairing and the antiferromagnetic channel. We plot the ratio
\begin{eqnarray}\label{eq:R}
R=\frac{\lambda_F^d}{\lambda_F^a}=\frac{2\bar h_d^2\bar m_a^2}{\bar h_a^2\bar m_d^2}
\end{eqnarray}
in Fig. \ref{fig:R} for different values of $\mu$. In these plots the ratio $R$  first increases due to the generation of a non-vanishing $\bar h_d$, and subsequently decreases since $\bar m_a^2$ approaches zero or small values. We see the tendency that the maximum of $R$ increases with increasing $\mu$.

Our present truncation does not remain valid for low $k$ if $|\mu/t|$ becomes large. The reason is that $A_a$ or $\bar\lambda_a$ reach zero before either $\bar m_a^2$ or $\bar m_d^2$ vanishes. This means that our approximation for the momentum dependence  in $P_a(Q)$, or for the polynomial approximation for $U_a$, becomes insufficient. An extended truncation can cure these problems, which will be addressed in a future publication. In our figures the breakdown of the approximations can be seen by the end of the curves in Fig. \ref{fig:phasediagronlyantif} and Fig. \ref{fig:R}. The insufficiency of the truncation is the reason why we cannot access the values of $|\mu|$ and $k$ where $R$ becomes large in the present paper.

\section{Conclusions}

We have shown how the renormalization flow turns the microscopic Hubbard model into an effective boson exchange model at intermediate length scales $~k^{-1}$. This ``mesoscopic'' model is characterized by the exchange of collective electron or electron-hole pairs, which are described by bosonic fields. The most important channels are the exchange of electron-hole pairs of the antiferromagnetic type -- the $\mathbf a$-boson -- and $d$-wave electron pairs -- the $d$-boson. The coupling of these bosons to the electrons is described by Yukawa type couplings $\bar h_a$, $\bar h_d$, which multiply the form factors appropriate for the respective channels. The effective model is characterized by the effective action, cf. Eqs. (\ref{eq:simplesttruncation}), (\ref{eq:pured}), (\ref{eq:Fd}), $\Gamma=\Gamma_F+\Gamma_a+\Gamma_{Fa}+\Gamma_d+\Gamma_{Fd}$. Besides the Yukawa couplings, important ingredients are the propagator and the effective potential for the $\mathbf a$-bosons and the $d$-bosons.

The effective boson exchange model has several coupling constants, as $\bar h_a$, $\bar h_d$, $\bar m_a^2$, $\bar m_d^2$, as well as bosonic self interactions and couplings parameterizing the bosonic propagators (cf. appendix). These couplings depend on the momentum scale $k$ at which the effective model is considered. We may consider the effective boson exchange model as a coarse grained version of the microscopic Hubbard model, with coarse graining length $k^{-1}$. Its couplings can therefore be related to the couplings $t$, $t'$ and $U$ of the Hubbard model, and we have done so by the use of the functional renormalization flow. This flow shows how an effective coupling in the $d$-wave pairing channel is generated by the fluctuations of the $\mathbf a$-bosons. 

One may also consider the effective boson exchange model with arbitrary parameters. This can then be interpreted as a generalization of the Hubbard model. The particular case $\bar h_a=0$, where the $\mathbf a$-bosons play no role, has already been discussed with the help of functional flow equations in Ref. \cite{kw07}.
For low temperatures one finds superfluidity. The investigation of \cite{kw07} has covered the critical behavior at the phase transition and the low temperature ``ordered phase''. The phase transition was found to be in the universality class of the Kosterlitz-Thouless phase transition.
Reinterpreted in the context of our investigation of the Hubbard model, the model \cite{kw07} becomes a valid approximation if at some scale $k$ the antiferromagnetic channel coupling $\lambda_F^a=-\bar h_a^2/2\bar m_a^2$ becomes much smaller than the $d$-wave coupling $\lambda_F^d=-\bar h_d^2/\bar m_d^2$. The corresponding large values of $R$ are reached, however, only in a region for $\mu/t$ where our present approximation for the propagator of the antiferromagnetic bosons becomes insufficient.

As we have mentioned in Sect. V, our present investigation is limited to the range of $k$ where the propagator for the $\mathbf a$-boson is well described by the approximation discussed in the appendix. In particular, this requires $A_a\geq 0$. We implicitly assume that commensurate antiferromagnetic fluctuations are dominant as compared to incommensurate antiferromagnetism or ferromagnetism. The latter play an important role in other parameter ranges of this model \cite{hlubina,arrachea,honerkampsalmhofer01,pandey,hankevych}. 
Within the present truncation the positivity of $A_a$ is not realized for all $\mu\,,t\,,t'\,,U\,,T$ if $k$ becomes small.
Also the simple dependence of $P_a$ on the Matsubara frequency does not remain appropriate for $T\rightarrow 0$. We will discuss a more general form of $P_a$, as well as a momentum dependent propagator for the $d$-boson, in a forthcoming work. We are confident that a suitable truncation of the functional renormalization group equations will give access to the whole phase diagram of the Hubbard model.
\\

{\bf Acknowledgments}: We thank S. Diehl, J. M. Pawlowski, H. Gies and P. Strack for useful discussions.  HCK acknowledges financial support by the DFG research unit FOR 723 under the contract WE 1056/9-1. JAM acknowledges financial support by the Helmholtz-University Young Investor Grant VH-NG-332.


\renewcommand{\thesection}{}
\renewcommand{\thesubsection}{A{subsection}}
\renewcommand{\theequation}{A \arabic{equation}}

\begin{appendix}
\section*{Appendix: Bosonic propagator}\setcounter{equation}{0}

\begin{figure}[t]
\includegraphics[width=70mm,angle=0.]{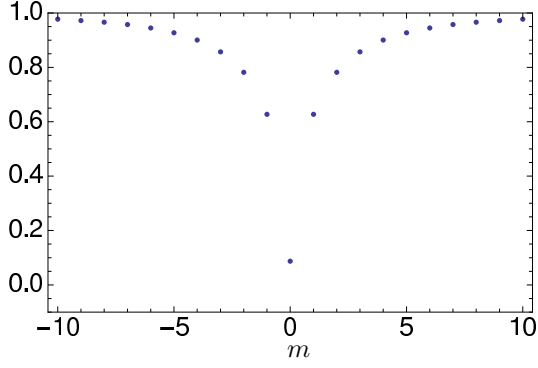}
\caption{\small{Mean field calculation of the frequency dependence of the  bosonic propagator at zero spatial momenta, $\bar m_a^2|_{\Lambda}+\Delta G^{-1}_a(2\pi mT,\mathbf q=0)$ for $U/t=3$, $T/t=0.25$, $t'=0$ and $\mu=0$.
}}
\label{fig:menafieldapropmfreq}
\end{figure}

\begin{figure}[b]
\includegraphics[width=85mm,angle=0.]{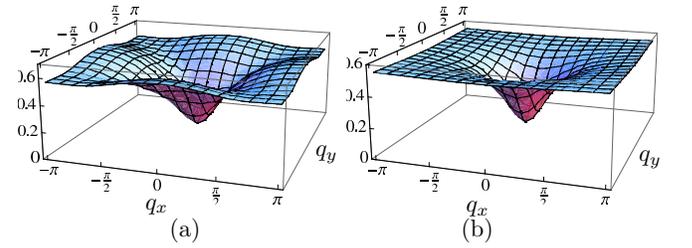}
\caption{\small{
The mean field approximation for the spatial momentum dependence of the bosonic propagator at zero frequency $\bar m_a^2|_{\Lambda}+\Delta G^{-1}_a(0,\mathbf{q})$ is shown in (a). We use $U/t=3$, $\mu=0$, $t'=0$ and $T/t=0.25$. In (b) the parameterization in terms of Eq. (\ref{eq:apropparam}) is displayed for a renormalization scale $k=0$.}}
\label{fig:menafieldapropmom}
\end{figure}

In order to gain some first information on the general shape of the momentum dependent piece in the inverse propagator for the antiferromagnetic boson, we compute the contribution from the fermionic loop
\begin{eqnarray}\label{eq:rhoferm}
\Delta G^{-1}_{a}(Q)=\sum_{P}
\frac{\bar{h}_{a}^2}
{P_F(Q+P+\Pi)P_F(P)}   +(Q\rightarrow -Q)
\,.\end{eqnarray}
The general features of this one loop contribution will be used in order to motivate a suitable truncation. Performing the Matsubara sum one finds 
\begin{eqnarray}\label{eq:apropkorr}
\Delta G^{-1}_{a}(\omega,\mathbf{q})&=&-\frac{\bar{h}_{a}^2}{2}
\int_{-\pi}^{\pi}\frac{d^2p}{(2\pi)^2}
\frac{\tanh(\frac{\xi_{\mathbf{p}}}{2T})-\tanh(\frac{\xi_{\mathbf{q+p+\pi}}}{2T}+\frac{i\omega}{2T})}{\xi_{\mathbf{p}}-\xi_{\mathbf{q+p+\pi}}-i\omega}\nonumber\\
& &\!\!+(Q\rightarrow -Q)
\,.\end{eqnarray}
In Fig. \ref{fig:menafieldapropmfreq} we plot the frequency dependence at vanishing spatial momenta while the dependence on spatial momenta at vanishing frequency is plotted in Fig. \ref{fig:menafieldapropmom} (a).

We approximate the inverse bosonic propagator at zero frequency by
\begin{eqnarray}\label{eq:apropparam}
P_{a,k}(0,\mathbf{q})=\Delta G^{-1}_{a,k}(Q)+\bar{m}^2_a|_{\Lambda}-\bar{m}^2_a|_k=A_a\frac{[\mathbf{q}]^2D^2}{D^2+[\mathbf{q}]^2}
\,,\end{eqnarray}
where $[\mathbf{q}]^2$ is defined as $[\mathbf{q}]^2=q_1^2+q_2^2$ for $q_i\in [-\pi,\pi]$ and continued periodically otherwise. It is convenient to define the gradient coefficient $A_a$ and the shape coefficient $D$ by
\begin{eqnarray}\label{eq:Aa}
A_a=\frac{1}{2}\frac{\partial^2}{\partial l^2}P_a(0,l,0)\big{|}_{l=0}
\end{eqnarray}
and 
\begin{eqnarray}\label{eq:D}
D^2=\frac{1}{A_a}\big(P_a(0,\pi,\pi)-P_a(0,0,0)\big)
\,.\end{eqnarray}
Comparison between Figs. \ref{fig:menafieldapropmom}a and \ref{fig:menafieldapropmom}b shows that this choice reproduces the mean field result rather well. (Note that $P_a(0,\pi,\pi)$ computed from Eqs. (\ref{eq:apropparam}) - (\ref{eq:D}) deviates slightly from the mean field result for this momentum choice.)
The lowest order perturbative contribution from the fermions reads
\begin{eqnarray}
A_a=\sum_Q\bar{h}_a^2\frac{\partial^2}{\partial l^2}\frac{1}{P^k_F(Q)P^k_F(K+Q+\Pi)}\Bigg{|}_{l=0}
\,.\end{eqnarray}

For frequencies different from zero we set for all $\mathbf q$
\begin{eqnarray}\label{eq:patunc}
P_a(\omega \neq 0,\mathbf{q})=\bar{m}^2_a|_{\Lambda}-\bar{m}^2_a|_k
\,.\end{eqnarray}
In practice, this means that we include the fluctuation effects only for the zero Matsubara frequency mode.
In the temperature range of interest in this work, this is a well justified approximation, cf. Fig. \ref{fig:menafieldapropmfreq}. On the other hand, the truncation (\ref{eq:patunc}) is not suited for very low T and therefore limits the temperature range for our investigation in the present paper.

Within the functional renormalization group approach, we describe the scale dependence of the inverse bosonic propagator by flow equations for the parameters $A_a$ and $D$. The flow equation for the gradient coefficient is
\begin{eqnarray}\label{eq:flowAa}
\partial_kA_a=\sum_Q\bar{h}_a^2\tilde\partial_k\frac{\partial^2}{\partial l^2}\frac{1}{P^k_F(Q)P^k_F(K+Q+\Pi)}\Bigg{|}_{l=0}
\,,\end{eqnarray}
where $K=(0,l,0)$. At the initial scale the gradient coefficient vanishes $A_a|_{\Lambda}=0$. The boson is not dynamic on this scale and can be regarded as an auxiliary field.
The difference $P_a(0,\pi,\pi)-P_a(0,0,0)$ in Eq. (\ref{eq:D}) is also scale dependent. From this, and Eqs. (\ref{eq:D}), (\ref{eq:flowAa}), one infers the flow equation for $D$. Again, $D$ vanishes at the initial scale $\Lambda$. 
The result of our approximation (\ref{eq:apropparam}), as calculated from the flow equation, is displayed in Fig. \ref{fig:menafieldapropmom} b.
Comparison with the one loop result in Fig. \ref{fig:menafieldapropmom} a shows satisfactory agreement.

The infrared cutoff $R_k^a$ for the $\mathbf a$-boson is adapted to Eq. (\ref{eq:apropparam}). We use an ``optimized cutoff'' \cite{litim1,litim2,PawlowskiBB}
\begin{eqnarray}
R^a_k(Q)=A_{a}\cdot(k^2-\frac{[\mathbf{q}]^2D^2}{D^2+[\mathbf{q}]^2})\Theta(k^2-\frac{[\mathbf{q}]^2D^2}{D^2+[\mathbf{q}]^2})
\,.\end{eqnarray}

\end{appendix}



\begin{thebibliography}{12}
\expandafter\ifx\csname natexlab\endcsname\relax\def\natexlab#1{#1}\fi
\expandafter\ifx\csname bibnamefont\endcsname\relax
  \def\bibnamefont#1{#1}\fi
\expandafter\ifx\csname bibfnamefont\endcsname\relax
  \def\bibfnamefont#1{#1}\fi
\expandafter\ifx\csname citenamefont\endcsname\relax
  \def\citenamefont#1{#1}\fi
\expandafter\ifx\csname url\endcsname\relax
  \def\url#1{\texttt{#1}}\fi
\expandafter\ifx\csname urlprefix\endcsname\relax\def\urlprefix{URL }\fi
\providecommand{\bibinfo}[2]{#2}
\providecommand{\eprint}[2][]{\url{#2}}

\bibitem{hubbard}
J. Hubbard,
  Proc. R. Soc. London, Ser. A \textbf{276}, 238 (1963).

\bibitem{kanamori}
J. Kanamori,
  Prog. Theor. Phys. \textbf{30}, 275 (1963).

\bibitem{gutzwiller}
M. C. Gutzwiller,
  Phys. Rev. Lett. \textbf{10}, 159 (1963).

\bibitem{anderson}
P. W. Anderson,
  Science \textbf{235}, 1196 (1987).

\bibitem{scalapino}
D. J. Scalapino,
  Physics Reports \textbf{250}, 329 (1995).

\bibitem{miyake}
K. Miyake, S. Schmitt-Rink and C. M. Varma,
  Phys.\ Rev. B \textbf{34}, 6554 (1986).

\bibitem{loh}
D. J. Scalapino, E. Loh and J. E. Hirsch,
  Phys.\ Rev. B \textbf{34}, 8190 (1986).

\bibitem{bickers}
N. E. Bickers, D. J. Scalapino and R. T. Scalettar,
  Int. J. Mod. Phys. B \textbf{1}, 687 (1987).

\bibitem{bickersscalapinowhite}
N. E. Bickers, D. J. Scalapino and S. R. White,
  Phys. Rev. Lett. \textbf{62}, 961 (1989).

\bibitem{dahm}
T. Dahm and L. Tewordt,
  Phys. Rev. Lett. \textbf{74}, 793 (1995).

\bibitem{schmalian}
J. Schmalian,
  Phys. Rev. Lett. \textbf{81}, 4232 (1998).

\bibitem{millismonien}
A. J. Millis, H. Monien and D. Pines,
  Phys. Rev. B \textbf{42}, 167 (1990).

\bibitem{millisbalatsky}
P. Monthoux, A. V. Balatsky and D. Pines,
  Phys. Rev. Lett. \textbf{67}, 3448 (1991).

\bibitem{moreo}
A. Moreo,
  Phys. Rev. B \textbf{45}, 5059 (1992).

\bibitem{assaad}
F. F. Assaad, W. Hanke and D. J. Scalapino,
  Phys. Rev. Lett. \textbf{71}, 1915 (1993).

\bibitem{dagotto}
E. Dagotto,
  Rev. Mod. Phys. \textbf{66}, 763 (1994).

\bibitem{maier}
T. A. Maier, M. Jarrell, T. C. Schulthess, P. R. C. Kent and J. B. White,
  Phys. Rev. Lett. \textbf{95}, 237001 (2005).

\bibitem{veilleux}
A. F. Veilleux, A.-M. Dar\'{e}, L. Chen, Y. M. Vilk and A.-M. S. Tremblay,
  Phys. Rev. B \textbf{52}, 16 255 (1995).

\bibitem{zhang}
S. Zhang, J. Carlson and J. E. Gubernatis,
  Phys. Rev. Lett. \textbf{78}, 4486 (1997).

\bibitem{aimi}
T. Aimi and M. Imada,
  J. Phys. Soc. Jpn. \textbf{76}, 113708 (2007).

\bibitem{zanchi1}
D. Zanchi and H. J. Schulz,
  Z.Phys. \textbf{B103}, 339 (1997).

\bibitem{zanchi2}
D. Zanchi and H. J. Schulz,
  Europhys. Lett. \textbf{44}, 235 (1998).

\bibitem{halbothmetzner}
C. J. Halboth and W. Metzner,
  Phys.\ Rev. Lett. \textbf{85}, 5162 (2000).

\bibitem{halbothmetzner2}
C. J. Halboth and W. Metzner,
  Phys.\ Rev. B \textbf{61}, 7364 (2000).

\bibitem{salmhofer}
M. Salmhofer and C. Honerkamp,
  Prog. Theor. Phys. \textbf{105}, 1 (2001).

\bibitem{honerkamp01}
C. Honerkamp, M. Salmhofer, N. Furukawa and T. M. Rice,
  Phys. Rev. B \textbf{63}, 035109 (2001).

\bibitem{honerkampsalmhofer01}
C. Honerkamp and M. Salmhofer,
  Phys. Rev. Lett. \textbf{87}, 187004 (2001).

\bibitem{honerkamp02}
C. Honerkamp, M. Salmhofer and T. M. Rice,
  Eur. Phys. J. B \textbf{27}, 127 (2002).

\bibitem{katanin}
A. A. Katanin and A. P. Kampf,
  Phys. Rev. B \textbf{68}, 195101 (2003).

\bibitem{metznerreissrohe}
W. Metzner, J. Reiss and D. Rohe,
  Phys. stat. sol. (b) \textbf{243}, 46 (2006).

\bibitem{reissrohemetzner}
J. Reiss, D. Rohe and W. Metzner,
  Phys. Rev. B \textbf{75}, 075110 (2007).

\bibitem{cw93}
C. Wetterich,
  Phys. Lett. B \textbf{301}, 90 (1993).

\bibitem{berges_review02}
J. Berges, N. Tetradis and C. Wetterich,
  Phys. Rep. \textbf{363}, 223 (2002).

\bibitem{GiesWett}
H. Gies and C. Wetterich,
  Phys. Rev. D \textbf{65}, 065001 (2002).

\bibitem{PawlowskiBB}
J.M. Pawlowski,
  Annals Phys. \textbf{322}, 2831 (2007).

\bibitem{bbw04}
T. Baier, E. Bick and C. Wetterich,
  Phys.\ Rev. B \textbf{70}, 125111 (2004).

\bibitem{bbw05}
T. Baier, E. Bick and C. Wetterich,
  Phys.\ Lett. B \textbf{605}, 144 (2005).

\bibitem{kw07}
H. C. Krahl and C. Wetterich,
  Phys. Lett. A \textbf{367}, 263 (2007).

\bibitem{kosterlitzthouless73}
J. M. Kosterlitz and D. J. Thouless,
  J. Phys. C. \textbf{6}, 1181 (1973).

\bibitem{hubbardtransf}
J. Hubbard,
  Phys. Rev. Lett. \textbf{3}, 77 (1959).

\bibitem{stratonovich}
R. L. Stratonovich,
  Soviet. Phys. Doklady \textbf{2}, 416 (1958).

\bibitem{zanchiself}
D. Zanchi,
  Europhys. Lett. \textbf{55}, 376 (2001).

\bibitem{honerkampself}
C. Honerkamp and M. Salmhofer,
  Phys. Rev. B \textbf{67}, 174504 (2003).

\bibitem{arracheaself}
L. Arrachea and D. Zanchi,
  Phys. Rev. B \textbf{71}, 064519 (2005).

\bibitem{bbw00}
T. Baier, E. Bick and C. Wetterich,
  Phys. Rev. B \textbf{62}, 15471 (2000).

\bibitem{jaeckelw03}
J. Jaeckel and C. Wetterich,
  Phys. Rev. D \textbf{68}, 025020 (2003).

\bibitem{kohnluttinger}
W. Kohn and J. M. Luttinger,
  Phys. Rev. Lett. \textbf{15}, 524 (1965).

\bibitem{hlubina}
R. Hlubina, S. Sorella and F. Guinea,
  Phys. Rev. Lett. \textbf{78}, 1343 (1997).

\bibitem{arrachea}
L. Arrachea,
  Phys. Rev. B \textbf{62}, 10 033 (2000).

\bibitem{pandey}
S. Pandey and A. Singh,
  Phys. Rev. B \textbf{75}, 064412 (2007).

\bibitem{hankevych}
V. Hankevych, B. Kyung and A.-M. S. Tremblay,
  Phys. Rev. B \textbf{68}, 214405 (2003).

\bibitem{litim1}
D. F. Litim,
  Phys. Lett. B \textbf{486}, 92 (2000).

\bibitem{litim2}
D. F. Litim,
  Phys. Rev. D \textbf{64}, 105007 (2001).

\end{thebibliography}
\end{document}